\documentclass[aps,prl,superscriptaddress,twocolumn]{revtex4}

\usepackage{graphicx} 
\usepackage{amsmath}
 
\newcommand{\beq}{\begin{equation}}
\newcommand{\enq}{\end{equation}}
\newcommand{\bea}{\begin{eqnarray}}
\newcommand{\ena}{\end{eqnarray}}

\newcommand{\ra}{\rangle}

\begin{document}

\title{Induced interactions and the superfluid transition temperature in a three-component Fermi gas}
\author{J.-P. Martikainen}
\affiliation{NORDITA, Roslagstullsbacken 21, 106 91 Stockholm, Sweden}
\author{J. J. Kinnunen}
\affiliation{Department of Applied Physics, Helsinki University of Technology, P.O. Box 5100,
02015 HUT, Finland}
\author{P. T\"{o}rm\"{a}}
\affiliation{Department of Applied Physics, Helsinki University of Technology, P.O. Box 5100,
02015 HUT, Finland}
\author{C. J. Pethick}
\affiliation{NORDITA, Roslagstullsbacken 21, 106 91 Stockholm, Sweden}
\affiliation{The Niels Bohr International Academy, The Niels Bohr Institute, Blegdamsvej 17, DK-2100
Copenhagen \O, Denmark}
\date{\today}

\begin{abstract}
We study many-body contributions 
to the effective interaction between fermions
in a three-component Fermi mixture. We find that effective interactions 
induced by the third component
can lead to a phase diagram different from that predicted if interactions with the third component are neglected. As a result, in a confining potential
a superfluid shell structure can arise even for equal populations of the components. 
We also find a critical temperature for the BCS transition in a
$^{6}{\rm Li}$ mixture which can deviate strongly from the 
one in a weakly interacting two-component system.

\end{abstract}
\maketitle

By using Feshbach resonances to change the effective 
interaction between ultracold atoms several groups
have probed the crossover from the Bardeen-Cooper-Schrieffer (BCS) superfluid
to a Bose-Einstein condensate 
of 
molecules~\cite{Jochim2003b,Regal2003a,Cubizolles2003a,Greiner2003a,Zwierlein2004a,Chin2004a,Kinast2004a,Zwierlein2005a,Zwierlein2006a,Partridge2006a}.
Studies of the crossover have provided important insights into fermionic superfluids
around the unitarity limit of strong interactions.
Importantly, it has become experimentally feasible to study also more complicated
mixtures than Fermi gases with two different atomic internal states. 
Bose-Fermi mixtures~\cite{Roati2002a,Ospelkaus2006a} and Bose-Einstein 
condensates with
many components have been created using many different
setups~\cite{Myatt1997a,Stenger1998a}. Also, heteronuclear Fermi-Fermi mixtures~\cite{Wille2008a,Spiegelhalder2009a}
and even heteronuclear Fermi-Fermi-Bose mixtures~\cite{Taglieber2008a} have
been recently demonstrated. 
In yet another breakthrough a three-component Fermi mixture 
of atoms in the three lowest hyperfine states of $^{6}{\rm Li}$
~\cite{Ottenstein2008a,Huckans2009a} has also been demonstrated.
Such multi-component 
systems have some intriguing similarities with quark matter counterparts
where color superconductivity may appear~\cite{Alford2008a}. 

The purpose of this Letter is to explore how
the induced interactions due to the third component modify the expected
behavior of three-component mixtures. This is important because, depending on parameters,
the many-body effects can change the effective interaction between
atoms substantially and on occasion even change the relative magnitudes
of couplings between different components. Such changes imply that
phase diagrams predicted using only two-body scattering properties 
can be incorrect.
Also, even when the corrections to the effective interactions are weak, they can cause
large changes to the critical temperature for the BCS transition. Indeed,
for a two component system Gorkov and Melik-Barkhudarov (GM) showed~\cite{Gorkov1961a}
that the perturbative correction 
to the effective interaction can reduce the critical 
temperature by a constant factor of $(4e)^{1/3}\approx 2.22$ in the weak-coupling limit.
Also in spin-density imbalanced systems such corrections have been shown to have a considerable effect~\cite{Gubbels2008a}. Here we will analyze the effects of analogous corrections on the three component system and
find important changes to the GM result.  As an interesting consequence of
these many-body corrections we predict in a spatially varying confining potential (typically 
harmonic trap) the appearance of superfluid shell structures even 
in the absence of population imbalance (polarization) of the components. These shell 
structures are due to {\it many-body effects only} and therefore fundamentally different
from earlier predictions of shell structures due to population, mass, or trapping potential
imbalance~\cite{Paananen2007a,Lin2006a}  
We also point out that many body effects due to the third component provide
a new way to tune the effective interaction between the two other fermions and that this
contribution can dominate over the usual GM contribution. 

Earlier, intriquing results have been found experimentally
for the critical temperature 
of iron-based multiband superconductors~\cite{Terashima2009a} and
degenerate three-component Fermi gases have been studied 
theoretically in a lattice~\cite{Honerkamp2004a,Rapp2008a}.
Furthermore, pairing~\cite{Paananen2006a,Paananen2007a,Bedaque2009a},
stability~\cite{Blume2008a}, and 
breached pairing~\cite{Errea2009a} have recently been studied
in a three-component fermionic mixtures.
However, these theoretical approaches did not consider situations directly relevant
to ongoing experiments and also did not study how the many-body effects
due to the presence of the third component influence the properties of the 
other two components. 
Some aspects of the many flavor problem 
were discussed by
Heiselberg {\it et al.}~\cite{Heiselberg2000a}.
\begin{figure}
\begin{tabular}{ll}
\includegraphics[width=0.50\columnwidth]{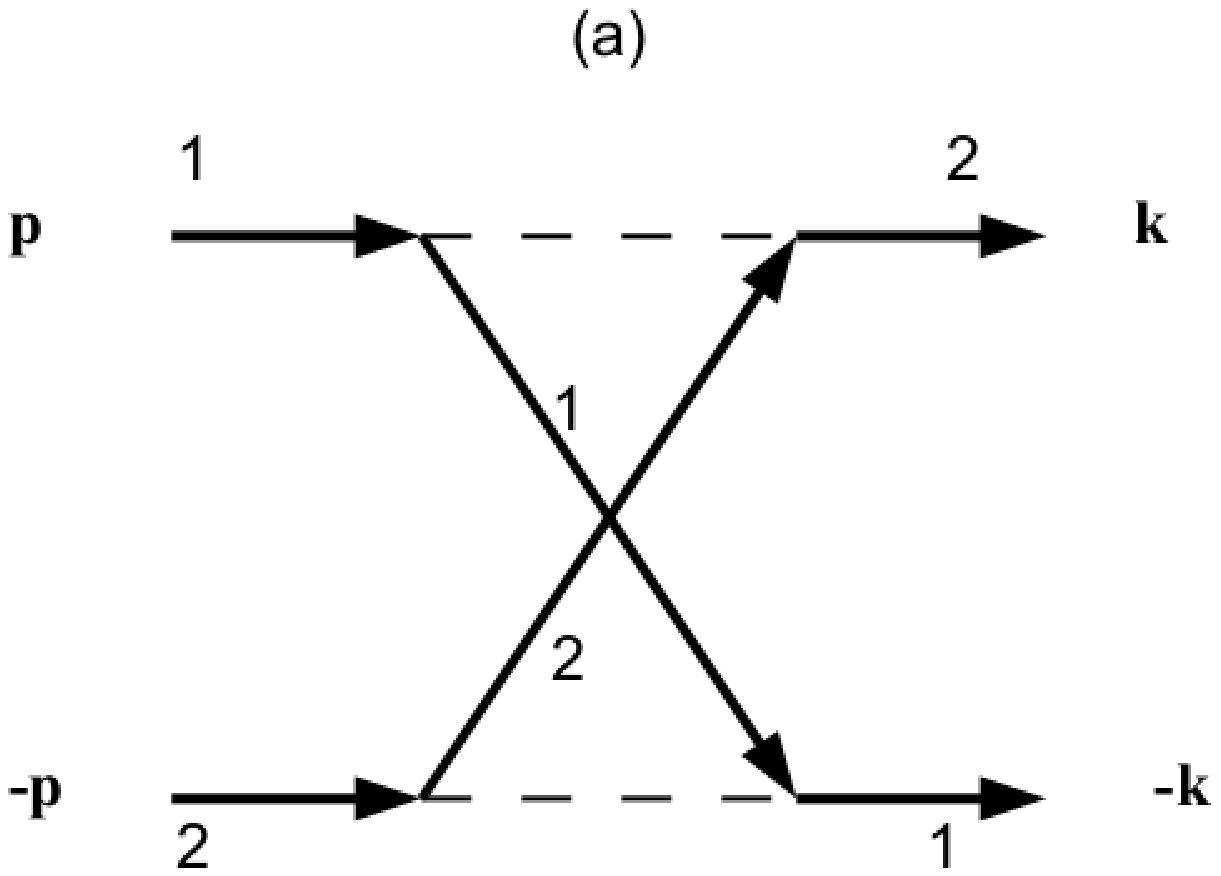} & \includegraphics[width=0.50\columnwidth]{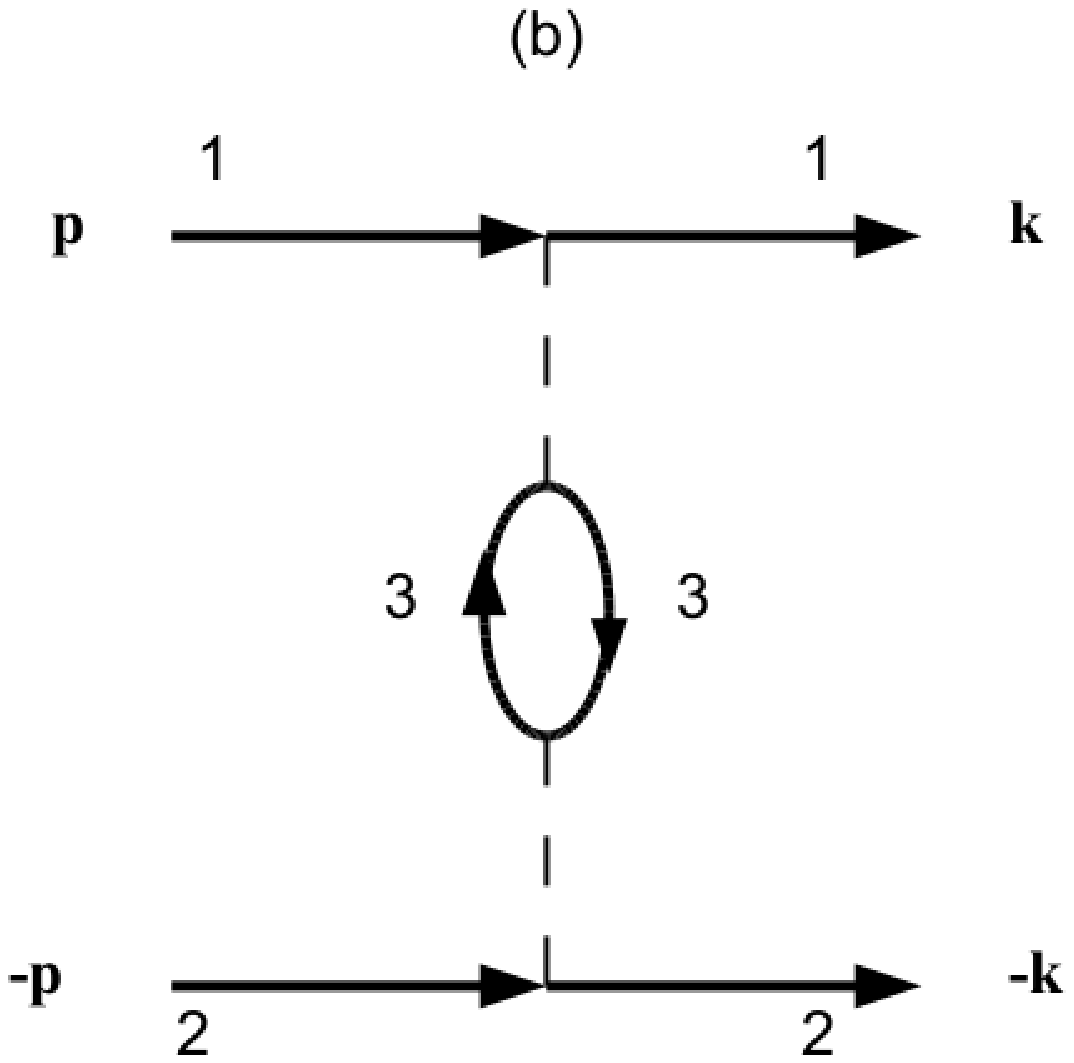}
\end{tabular}
\caption[Fig1]{Diagrams of second order in the interactions
for the induced interaction
between components $1$ and $2$. Solid lines represent atoms and the dashed lines
interactions between them, and numbers indicate the fermionic component.
The interactions are taken to be antisymmetrized with respect to interchange of spins and consequently interactions between atoms of the same species are absent. }
\label{fig:Diagrams}
\end{figure}

The relevant second order diagrams which give rise to induced interactions
between fermions of type $1$ and $2$ 
are shown in Fig.~\ref{fig:Diagrams}. In these diagrams the arrows
are the component propagators and dashed line is a contact interaction
with strength $U_{\alpha\beta}$ between components labeled by $\alpha,\beta\in \{1,2,3\}$ 
($\alpha\neq \beta$).
These couplings can be expressed in terms of the scattering lengths $a_{\alpha\beta}$
through $U_{\alpha\beta}=2\pi\hbar^2 a_{\alpha\beta}/m_{\alpha \beta}$, where $m_{\alpha \beta}=(1/m_\alpha+1/m_\beta)^{-1}$
is the reduced mass. Of the diagrams shown,
the diagram (a) is relevant in the case of a two component system with a contact 
interaction between unlike fermions~\cite{Baranov2008a} and 
for equal mass fermions gives rise to 
the GM correction mentioned
earlier. In a three-component system the diagram (b) describes the induced 
effect of the third component. 
Similar loop diagrams with the mediating fermion in component $1$ or $2$ are
forbidden in the s-wave scattering channel for symmetry reasons.
More formally the diagram (a) indicates the induced interaction
\beq
V^{G}({\bf p},{\bf p'})=
-U_{12}^2\sum_{\bf k} \frac{f\left[\xi_1({\bf k}+{\bf q}/2)\right]
-f\left[\xi_2({\bf k}-{\bf q}/2)\right]}
{\xi_1({\bf k}+{\bf q}/2)-\xi_2({\bf k}-{\bf q}/2)},\nonumber
\enq
where ${\bf q}={\bf p}+{\bf p'}$ and (b) describes the
induced interaction
\beq
V^{3c}({\bf p},{\bf p'})=
U_{13}U_{23}
\sum_{\bf k} 
\frac{f\left[\xi_3({\bf k}+{\bf q'}/2)\right]
-f\left[\xi_3({\bf k}-{\bf q'}/2)\right]}
{\xi_3({\bf k}+{\bf q'}/2)-\xi_3({\bf k}-{\bf q'}/2)},\nonumber
\enq
with ${\bf q'}={\bf p}-{\bf p'}$. In these formulas 
$\xi_\alpha({\bf k})=\hbar^2k^2/2m_\alpha-\mu_\alpha$ are the free atom 
dispersion relations
and $f(\epsilon)$ is the Fermi distribution.

In the weak coupling limit the scattering processes around the Fermi 
surfaces dominate
and to find the effective coupling the induced interactions
are averaged over the Fermi surfaces. 
In this way we find that the effective coupling between
fermions of types $1$ and $2$ becomes
\beq
\begin{split}
&U_{12}^{\rm eff}=\frac{4\pi\hbar^2 a_{12}}{m_1}\left\{1+\frac{2}{\pi}\left[a_{12}k_{F,1}F\left(1\right)
\right.\right.\\
&\left.\left.
-\frac{a_{13}a_{23}}{a_{12}}\frac{(m_3+m_1)^2}{4m_1m_3}
\left(\frac{k_{F,3}^3}{k_{F,1}^2}\right)F\left(\frac{k_{F,1}}{k_{F,3}}\right)
\right]\right\},
\label{eq:Ueff}
\end{split}
\enq
where $k_{F,\alpha}$ is the Fermi wavevector for the component $\alpha$ and 
we have assumed that fermions $1$ and $2$ both have a mass $m_1$ while
the third component has a mass $m_3$. 
The function $F(y)$ is given 
by the integral
\begin{equation}
F(y)=\int_0^{y} dw 
2w\left[\frac{1}{2}+\frac{(1-w^2)}{4w}\ln\left(\frac{|1+w|}{|1-w|}\right)\right],
\end{equation}
whose analytical solution is given by 
\begin{equation}
F(y)= 
\frac{1}{6} \left[-y \left(y^2-3\right) \log\left|\frac{y+1}{y-1}\right|
+2 \left(y^2+\log\left|y^2-1\right|\right)\right]
\nonumber
\end{equation}
The effective interactions in other channels can be found in the same way.

In Eq.~(\ref{eq:Ueff}) the first term describes the two-body scattering
in the absence of Fermi seas, the second term gives rise
to the GM correction, and the third term describes the effect of the 
interactions with the third component and its Fermi sea.
The correction due to the second term 
always suppresses the critical temperature
for the BCS transition. However, the last term is proportional to the product
$a_{23}a_{13}$ and can have either 
sign. Therefore,
the presence of the third component can
either suppress or enhance the critical temperature.

\begin{figure}
\begin{tabular}{ll}
\includegraphics[width=0.75\columnwidth]{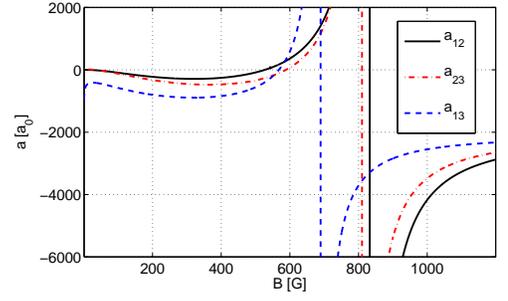}
\end{tabular}
\caption[Fig2]{Scattering lengths in units of the Bohr radius between 
$^{6}{\rm Li}$ atoms as a function of magnetic field. (The figure
is taken from Ref.~\cite{Julienneprivatecomm2009}.)
}
\label{fig:Li6scatteringlengths}
\end{figure}

Since three component systems have 
been demonstrated using $^{6}{\rm Li}$ atoms, 
let us now investigate these many-body effects using the 
coupled channel scattering data
for $^{6}{\rm Li}$~\cite{Julienneprivatecomm2009}.  
In Fig.~\ref{fig:Li6scatteringlengths} we show the scattering lengths
bet\-ween different components of the $^{6}{\rm Li}$ mixture
($|1\ra$, $|2\ra$, $|3\ra$ refer to the states
$|F,m_F\ra=|1/2,1/2\ra$, $|1/2,-1/2\ra$, $|3/2,-3/2\ra$, respectively).
It can be seen that, in the absence of many-body corrections, the $1-3$ channel
has the most negative scattering length for weaker magnetic fields, while
at magnetic fields above the Feshbach resonances the $1-2$ channel eventually becomes
dominant. In the simple mean-field picture one would infer that these channels
are also the ones with highest critical temperatures. However, when we include
induced interactions, density dependencies appear in the effective coupling strengths
and change the simple picture in which the third component is neglected.

\begin{figure}
\begin{tabular}{ll}
\includegraphics[width=0.5\columnwidth]{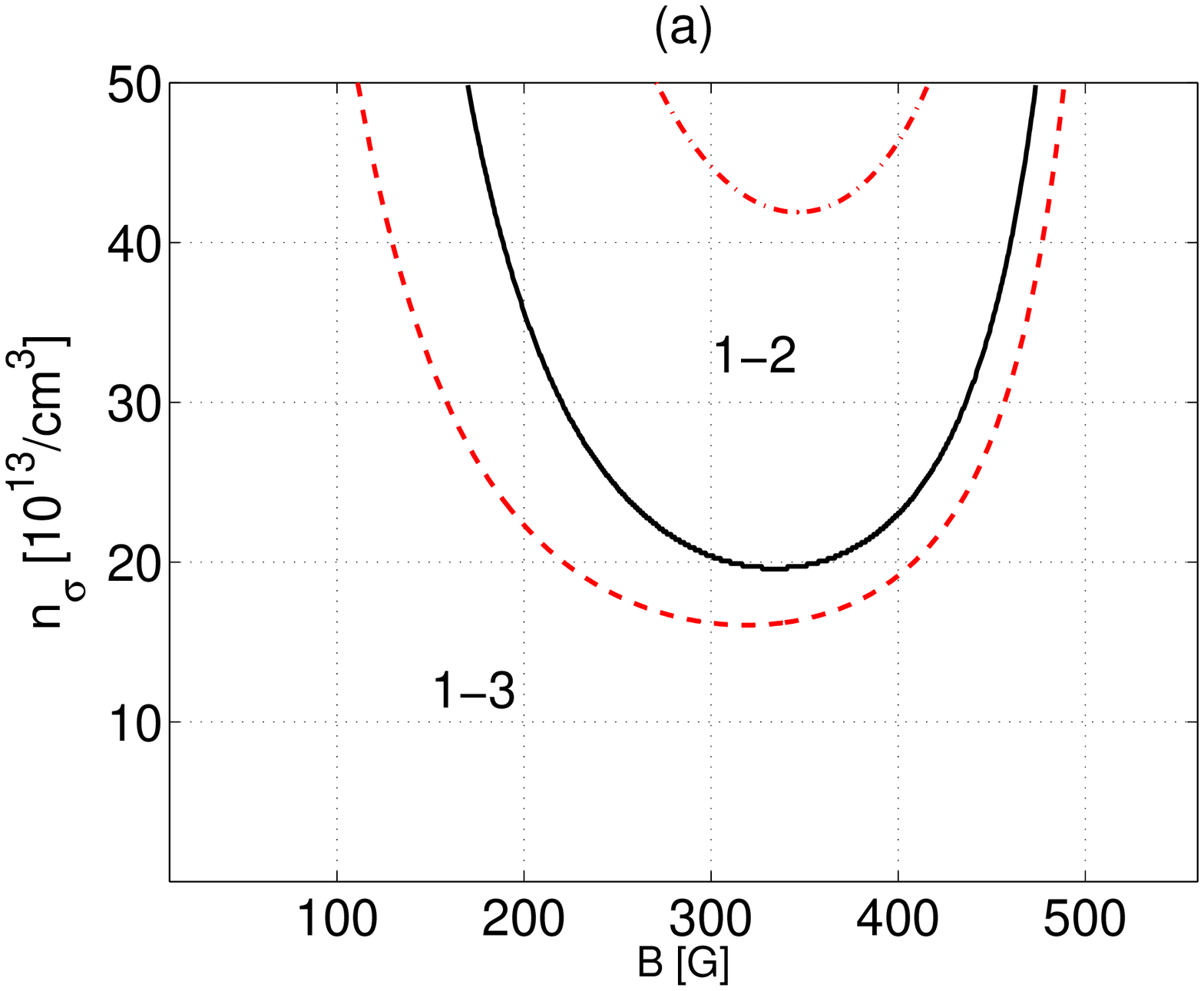} &
 \includegraphics[width=0.468\columnwidth]{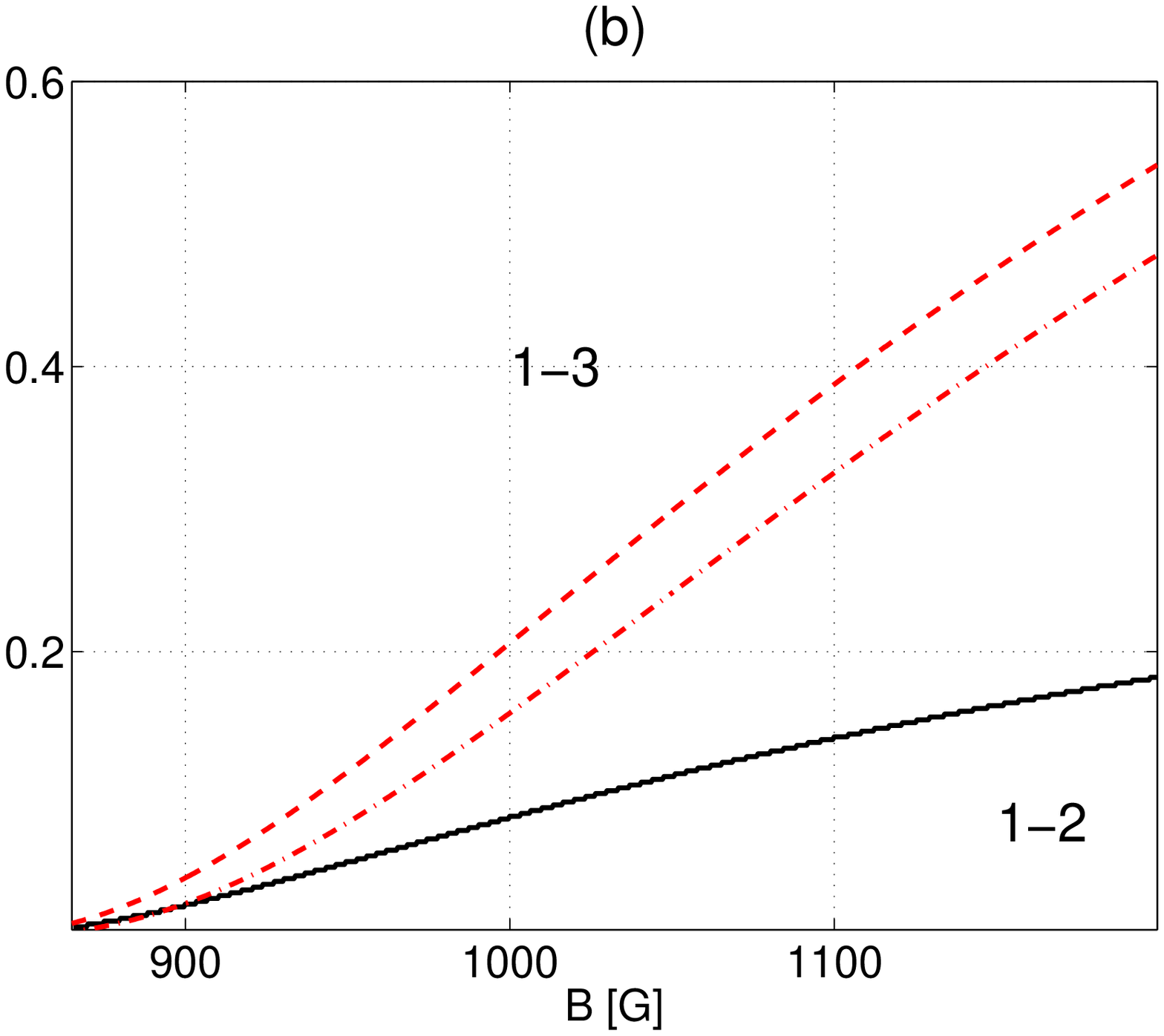}
\end{tabular}
\caption[Fig3]{The dominant interaction channel
including induced interactions for an equal-density $^{6}{\rm Li}$ mixture.
The solid black lines indicate phase boundaries.
Below the (red) dashed line
both $|k_Fa_{13}|$ and $|k_F^2a_{23}a_{12}|$ are less than one, while
below the (red) dot-dashed line both $|k_Fa_{12}|$ and $|k_F^2a_{23}a_{13}|$ 
are less than one. In (a) we show the interesting regions 
at low magnetic fields and in (b) the behavior at higher magnetic fields.
}
\label{fig:phasediagramLi6}
\end{figure}

Let us first explore the case where all $^{6}{\rm Li}$ components
have the same density.
In Fig.~\ref{fig:phasediagramLi6} we show the domi\-nant coupling channel
in  the magnetic field--density plane.
It can be seen that below the Feshbach resonance the $1-3$ channel
dominates for smaller densities, but for densities higher than 
about $2\cdot 10^{14}/{\rm cm^3}$ there is a possibility that the
$1-2$ channel becomes dominant. 
At higher magnetic fields we find a possibility of dominant
$1-3$ coupling in the region where the scattering lengths would predict
the $1-2$ channel. In experiments the atoms are trapped, 
and applying a local density approximation (which has been sufficient
to describe many experiments) with our results
suggests the interesting possibility of different superfluid phases appearing
in different parts of the cloud even in the absence of polarization
or unequal trapping potentials/masses~\cite{Paananen2007a}. This
possibility is a many-body effect caused by the induced interactions only;
for a balanced system at zero temperature, simple mean field theory would
not predict the shell structures that arise from the density dependence
of the GM correction as shown here.

\begin{figure}
\begin{tabular}{ll}
\includegraphics[width=0.95\columnwidth]{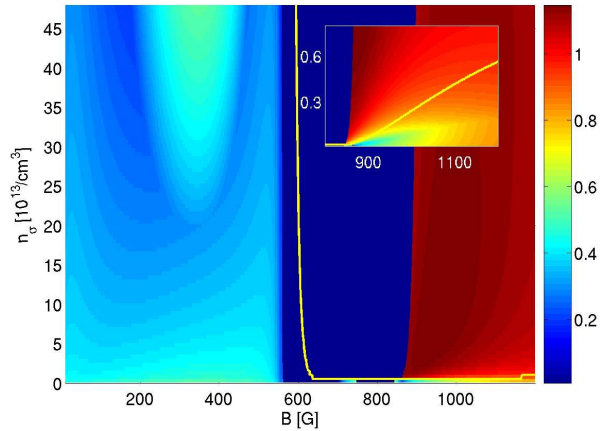}
\end{tabular}
\caption[Fig4]{The highest 
fraction $T_c/T_{c,0}$ of the critical temperatures with and
without induced effects for the equal-density $^{6}{\rm Li}$
mixture. 
The fraction is only computed in the regions where the dominant effective 
interaction is attractive and set to zero elsewhere.
Below the yellow
line $|k_Fa^{\rm eff}|<1$. Regions where the optimal pairing channel is changed
are visible as kinks in the fraction $T_c/T_{c,0}$. 
The inset shows a close-up into the region
of high magnetic fields and low densities.
}
\label{fig:Tccorrections}
\end{figure}

It is important to investigate what these results 
 imply for the critical tempera\-ture.
In attractive dilute Fermi gases,  
the critical temperature for the BCS transition in the weak coupling limit is 
$k_B T_c/\epsilon_F \propto  \exp[-\pi/(2k_F|a^{\rm eff}|)]$ 
where $a^{\rm eff}=U^{\rm eff}N(\epsilon_F)\pi/(2k_F)$ is the effective scattering length
and $N(\epsilon_F)$ is the density of states at the Fermi level.
We now use this result to estimate the many-body correction to $T_c$
in a three-component $^{6}{\rm Li}$ system. For simpli\-city we use the
above functional dependence in all regions where the coupling is attractive, but
indicate the regions where $|k_Fa|>1$ in the figures. In those regions the
weak-coupling formula is only suggestive. In Fig.~\ref{fig:Tccorrections}
we show the fraction $T_c/T_{c,0}$ 
for the equal density $^{6}{\rm Li}$ mixture ($T_{c,0}$ is the critical 
temperature in the absence of induced interactions). As can be seen, the correction to $T_c$ is 
often very different from the $1/2.22\approx 0.45$ GM result and shows a non-trivial
behavior as a function of the magnetic field and density due to complicated variation
of the scattering lengths. This also makes it possible that the pairing channel
is changed due to many-body effects. At high fields above the Feshbach resonances
it is possible that the critical temperature is enhanced since the effective scattering length
there becomes more negative due to induced interactions. However, this happens
in the region of stronger interactions where our results are not necessarily 
quantitatively accurate.

\begin{figure}
\begin{tabular}{ll}
\includegraphics[width=0.5\columnwidth]{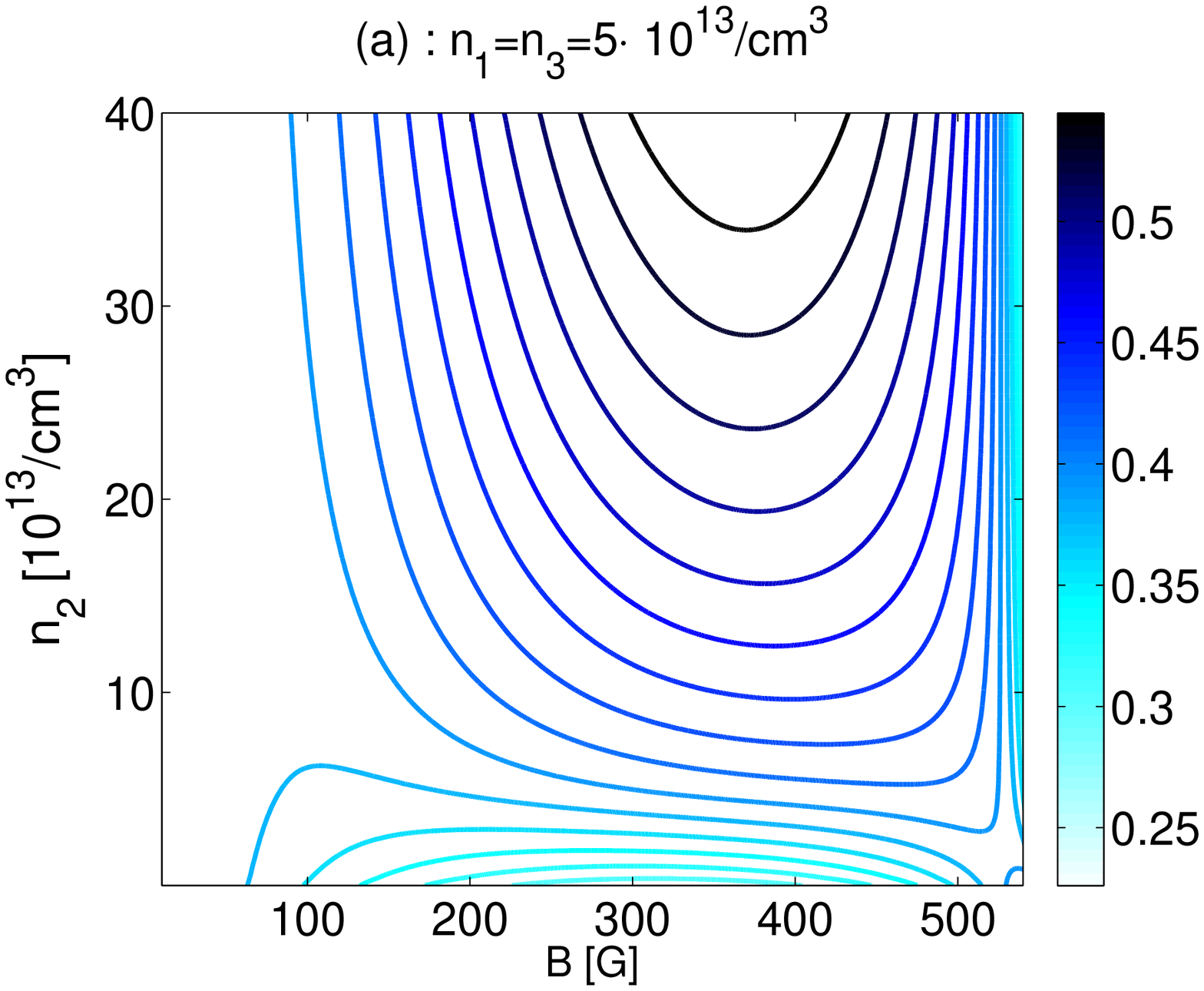} &
 \includegraphics[width=0.5\columnwidth]{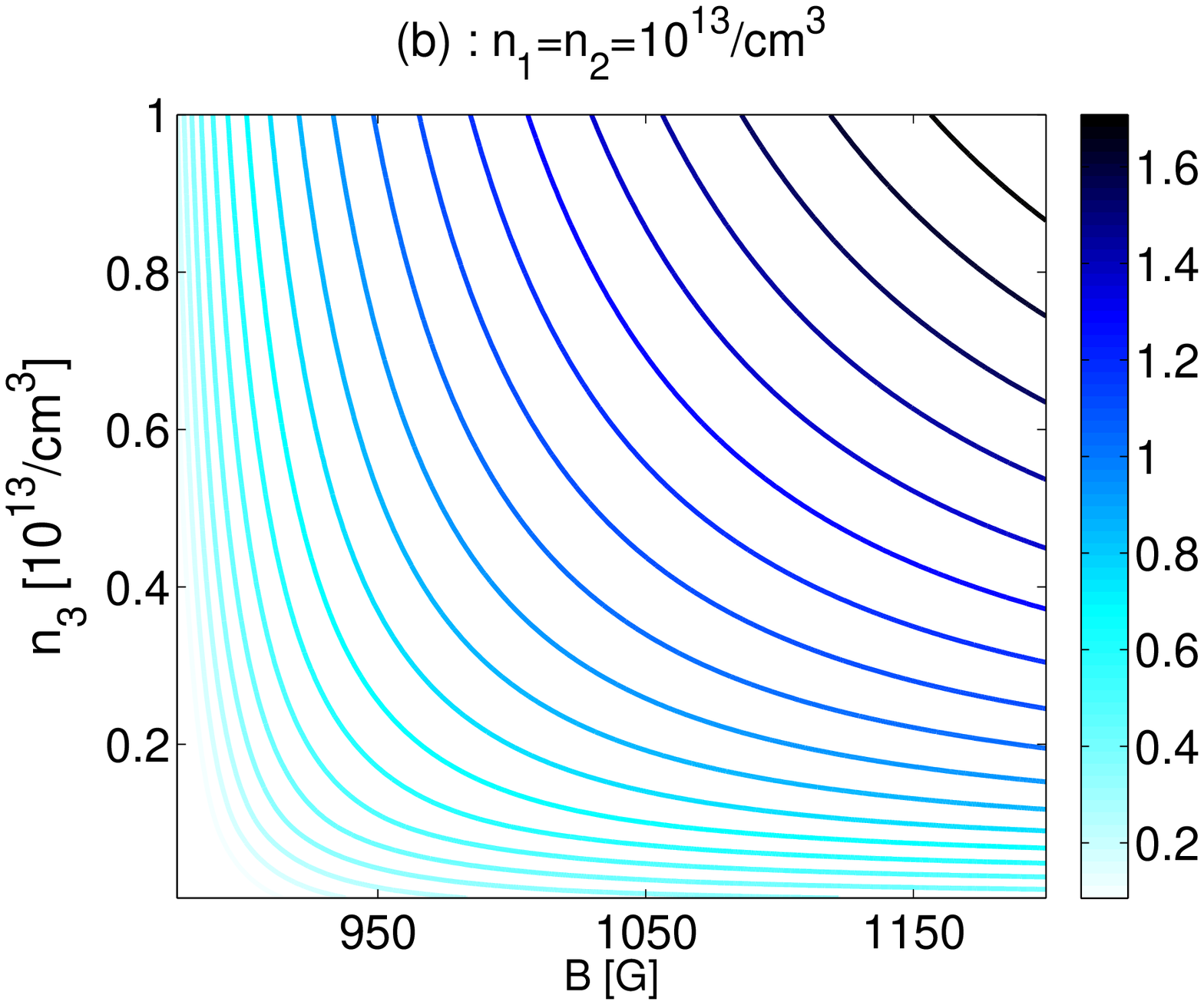}
\end{tabular}
\caption[Fig5]{The fraction $T_c/T_{c,0}$ of the critical temperatures with and
without induced effects for the $^{6}{\rm Li}$ mixture. In (a) we show
an example of how the critical temperature is changed in the $1-3$ channel when the 
density of the component $2$ is varied while in (b) we show the same
for the $1-2$ channel when the density of the component $3$ is varied.
We only focus on those magnetic fields where the pairing channel in question
is the dominant one.
}
\label{fig:Usentotune}
\end{figure}

In Fig.~\ref{fig:Usentotune} we demonstrate  another possibility 
for changing the critical temperature: the use of density imbalance.
In Fig.~\ref{fig:Usentotune} (a) we show an example of how $T_c$ in the $1-3$ channel is changed
as the density of the component $2$ is varied. It is again clear that the result deviates
substantially from the GM prediction, but the $T_c$ is nevertheless suppressed by the
component not involved in pairing. In Fig.~\ref{fig:Usentotune} (b) we show the similar
result in the $1-2$ channel which dominates at higher magnetic fields.
Due to the different behavior of the scattering lengths, here
the induced interactions can act
to enhance $T_c$ above the value predicted by the usual mean-field theory.

Finally, since heteronuclear fermionic mixtures are experimentally
feasible~\cite{Wille2008a,Taglieber2008a} let us briefly discuss what our results 
imply in that case.
A mass imbalance can be realized if the third component
is a different isotope, but also if the third component experiences
an optical lattice which changes its effective mass. In the latter case,
for the formulas
derived here to be valid, the filling fraction of all the components
should be much less than one.  For higher filling fractions the Fermi 
surface is no longer spherical and
the result would change considerably~\cite{Kim2009a}. We focus on a scenario
with equal masses for atoms of type $1$ and $2$,
since it is known that unequal mass of the interacting fermions suppresses 
the critical temperature~\cite{Baranov2008a,Paananen2007a} 
and for this reason the equal mass superfluidity appears more generic.
The effective interaction
between $1$ and $2$ is given by Eq.~(\ref{eq:Ueff}).
Note that since  
$(m_3+m_{1})^2/4m_{1}m_3>1$, induced interactions
become relatively stronger in an unequal mass mixture and
mass imbalance can be used to enhance the role of
many-body corrections. If 
$b=\left(a_{12}^2/|a_{13}a_{23}|\right)\left(k_{F,1}/k_{F,3}\right)^3F(1)/F(k_{F,1}/k_{F,3})$
and $b>1$, 
the contribution of the third component 
to the induced interaction becomes larger than the GM contribution
when $m_3/m_1> (2b-1)+ 2\sqrt{b(b-1)}$ or when $m_3/m_1<(2b-1)-2\sqrt{b(b-1)}$.
If $b<1$ the contribution due to the third component is dominant
for all mass ratios. However, for mass ratios larger than
about $13.6$ other physics can come into play since then
weakly bound diatomic molecules might become collisionally unstable~\cite{Petrov2005a}.

In this Letter we have explored the induced interactions and their role
in the BCS pairing in a three-component Fermi mixture. We found 
striking differences from physics ignoring these many-body corrections.
In particular, we found that when the induced interactions are taken into account,
the phase-diagram can change  drastically,
that shell structures in traps can appear even without number,mass, or trap imbalance, and
that the critical temperature for the BCS transition is strongly dependent
on the induced interactions in the three-component systems.


We thank Academy of Finland (Projects No. 213362, 217041,
217043, and 210953).

\end{document}